# Retrocausal Effects As A Consequence of Orthodox Quantum Mechanics Refined To Accommodate The Principle Of Sufficient Reason.

HENRY P. STAPP
LAWRENCE BERKELEY NATIONAL LABORATORY
UNIVERSITY OF CALIFORNIA
BERKELEY, CALIFORNIA 94720
JULY 18, 2011

**Abstract.** The principle of sufficient reason asserts that anything that happens does so for a reason: no definite state of affairs can come into being unless there is a sufficient reason why that particular thing should happen. This principle is usually attributed to Leibniz, although the first recorded Western philosopher to use it was Anaximander of Miletus. The demand that nature be rational, in the sense that it be compatible with the principle of sufficient reason, conflicts with a basic feature of contemporary orthodox physical theory, namely the notion that nature's response to the probing action of an observer is determined by pure chance, and hence on the basis of absolutely no reason at all. This appeal to pure chance can be deemed to have no rational fundamental place in reason-based Western science. It is argued here, on the basis of the other basic principles of quantum physics, that in a world that conforms to the principle of sufficient reason, the usual quantum statistical rules will naturally emerge at the pragmatic level, in cases where the reason behind nature's choice of response is unknown, but that the usual statistics can become biased in an empirically manifest and effectively retrocausal way when the reason for the choice is empirically identifiable. It is shown here that if the statistical laws of quantum mechanics were to be biased in this way then the basically forward-in-time unfolding of empirical reality described by orthodox quantum mechanics would generate the appearances of backward-time-effects of the kind that have been reported in the scientific literature.

**Keyword**s: Reason, Retrocausation, Orthodox Quantum Mechanics,

**This work was supported by the Director, Office of Science, Office of High Energy and Nuclear Physics, of the U.S. Department of Energy under contract DE-AC02-05CH11231**

## INTRODUCTION

An article recently published by the Cornell psychologist Daryl J. Bem [1] in a distinguished psychology journal has provoked a heated discussion in the New York Times. Among the discussants was Douglas Hofstadter who wrote that: "If any of his claims were true, then all of the bases underlying contemporary science would be toppled, and we would have to rethink everything about the nature of the universe."

It is, I believe, an exaggeration to say that if any of Bem's claims were true then "*all* of the bases underlying contemporary science would be toppled" and that "we would have to rethink *everything* about the nature of the universe". In fact, all that is required is a relatively small change in the rules, and one that seems reasonable and natural in its own right. The major part of the required rethinking was done already by the founders of quantum mechanics, and cast in more rigorous form by John von Neumann [2], more than eighty years ago.

According to the precepts of classical mechanics, once the physically described universe is created, it evolves in a deterministic manner that is completely fixed by mathematical laws that depend only on the present, or previously determined, values of evolving physically described properties. There are no inputs to the dynamics that go beyond what is specified by those physically described properties. [Here *physically described properties* are properties that are specified by assigning mathematical properties to space-time points, or to very tiny regions.] The increasing knowledge of human and other biological agents enters only as an *output* of the physically described evolution of the universe, and even nature itself is not allowed to interfere with the algorithmically determined mechanistic evolution.

This one-way causation from the physical to the empirical/epistemological has always been puzzling: Why should "knowledge" exist at all if cannot influence anything physical, and hence be of no use to the organisms that possess it. And how can something like an "idea", seemingly so different from physical matter, as matter is conceived of in classical mechanics, be created by, or simply *be*, the motion of physical matter?

But the basic precepts of classical mechanics are now known to be fundamentally incorrect: they cannot be reconciled with a plenitude of empirical facts discovered and verified during the twentieth century. Thus there is no reason to demand or believe that those puzzling properties of the classically conceived world must carry over to the real world, which conforms far better to the radically different precepts of quantum mechanics.

The founders of quantum theory conceived the theory to be a mathematical procedure for making practical predictions about future empirical-experiential findings on the basis of our present knowledge. According to this idea, quantum theory is basically about the evolution of knowledge. This profound shift is proclaimed by Heisenberg's assertion [3] that the quantum mathematics "represents no longer the behavior of the elementary particles but rather our knowledge of this behavior", and by Bohr's statement [4] that "Strictly speaking, the mathematical formalism of quantum mechanics merely offers rules of calculation for the deduction of expectation about observations obtained under conditions defined by classical physics concepts."

The essential need to bring "observations" into the theoretical structure arises from the fact that evolution via the Schroedinger equation, which is the quantum analog of the classical equations of motion, produces in general not a single evolving physical world that is compatible with human experience and observations, but rather a mathematical structure that corresponds to an increasingly smeared out mixture of many such worlds. Consequently, some additional process, beyond the one generated by Schroedinger equation, is needed to specify what the connection is between empirical/experiential findings and the physically described quantum state of the universe. Epistemological factors become thereby intertwined with the mathematically described physical aspects of the quantum mechanical conception of nature.

The founders of quantum mechanics achieved an important advance in our understanding of nature when they recognized that the mathematically-physically described universe that appears in our best physical theory represents not the world of material substance contemplated in the classical physics of Isaac Newton and his direct successors, but rather a world of *potentialities* or *possibilities* for our future acquisitions of *knowledge*. It is not surprising that a scientific theory designed to allow us to predict correlations between our shared empirical findings should incorporate, as orthodox quantum mechanics does: 1), a natural place for *"our knowledge"*, which is both all that is really known to us, and also the empirical foundation upon which science is based; 2), an account of the process by means of which we acquire our conscious *knowledge* of certain physically described aspects of nature; and 3), a statistical description, at the pragmatic level, of relationships between various features of the growing aspect of nature that constitutes "our knowledge". What is perhaps surprising is the ready acceptance by most western-oriented scientists and philosophers of the notion that the element of chance that enters quite reasonably into the *pragmatic* formulation of physical theory, in a *practical* context where many pertinent things may be unknown to us, stems from an occurrence of raw pure chance at the underlying *ontological* level. Ascribing such capriciousness to nature herself would seem to contradict the rationalist ideals of Western Science. From a strictly rational point of view, it not unreasonable to examine the mathematical impact of accepting, *at the basic ontological level*, Einstein's dictum that: "God does not play dice with the universe", and to attribute the effective entry of pure chance at the pragmatic level to our lack of knowledge of the *reasons* for the "choices on the part of nature" to be what they turn out to be.

These "random" quantum choices are key elements of orthodox quantum mechanics, and the origin of these choices is therefore a fundamental issue. Are they really purely random, as contemporary orthodox theory asserts? Or could they stem at the basic ontological level from sufficient reasons?

It is well known---as will be reviewed presently---that biasing the weights of the random quantum choices, relative to the weights prescribed by orthodox quantum theory, leads to an apparent breakdown of the normal causal structure of phenomena. This breakdown of the causal structure dovetails neatly with the empirical findings reported by Bem, and the similar retrocausal findings reported earlier by others [5,6]. In particular, the rejection of the intrinsically "irrational" idea that definite choices can pop out of nothing at all, and the acceptance, instead, of the principle of sufficient reason, yields a rational revision of orthodox quantum mechanics that can naturally accommodate the reported retrocausal phenomena, while preserving most of orthodox quantum mechanics. This revision allows nature's choices to provide more high-level guidance to the evolution of the universe than the known-to-be-false precepts of classical mechanics allow.

## IMPLEMENTING THE PRINCIPLE OF SUFFICIENT REASON

I make no judgment on the significance of the purported evidence for the existence of various retrocausal phenomena. That I leave to the collective eventual wisdom of the scientific community. I am concerned here rather with essentially logical and mathematical issues, as they relate to the apparent view of some commentators that scholarly articles reporting the existence of retrocausal phenomena should be banned from the scientific literature, essentially for the reason articulated in the New York Times by Douglas Hofstadter, namely that the actual existence of such phenomena is irreconcilable with what we now (think we) know about the structure of the universe; that the actual existence of such phenomena would require a wholesale abandonment of basic ideas of contemporary physics. That assessment is certainly not valid, as will be shown here. Only a limited, and intrinsically reasonable, modification of the existing orthodox QM is needed in order to accommodate the reported data.

In order for science to be able to confront effectively purported phenomena that violate the prevailing basic theory what is needed is an alternative theory that retains the valid predictions of the currently prevailing theory, yet accommodates in a rationally coherent way the purported new phenomena.

If the example of the transition from classical physics to quantum physics can serve as an illustration, in that case we had a beautiful theory that had worked well for 200 years, but that was incompatible with the new data made available by advances in technology. However, a new theory was devised that was closely connected to the old one, and that allowed us to recapture the old results in the appropriate special cases, where the effects of the nonzero value of Planck's constant could be ignored. The old formalism was by-and-large retained, but readjusted to accommodate the fact that pq-qp was non-zero. Yet there was also *a rejection of a basic classical presupposition*, namely the idea that a physical theory should properly be exclusively about connections between physically described material events. The founders of quantum theory insisted [7] that their physical theory was a pragmatic theory --- i.e., was directed at predicting practically useful connections between empirical (i.e., experienced) events.

This original pragmatic Copenhagen QM was not suited to be an ontological theory, because of the movable boundary between the aspects of nature described in *classical* physical terms and those described in *quantum* physical terms. It is certainly not ontologically realistic to believe that the pointers on observed measuring devices are built out of *classically* conceivable electrons and atoms, etc. The measuring devices, and also the bodies and brains of human observers, must be understood to be built out of quantum mechanically described particles. That is what allows us to understand and describe many observed properties of these physically described systems, such as their rigidity and electrical conductance.

Von Neumann's analysis of the measurement problem allowed the quantum state of the universe to describe the entire physically described universe: everything that we naturally conceive to be built out of atomic constituents and the fields that they generate. This quantum state is described by assigning mathematical properties to space-time points (or tiny regions). We have a deterministic law, the Schroedinger equation, that specifies the mindless, essentially mechanical, evolution of this quantum state. But this quantum mechanical law of motion generates *a huge continuous smear of worlds of the kind that we actually experience.* For example, as Einstein emphasized, the position of the pointer on a device that is supposed to tell us the *time* of the detection of a particle produced by the decay of a radioactive nucleus, evolves, under the control of the Schroedinger equation, into a *continuous smear of positions corresponding to all the different possible times of detection*; not to a single position, which is what we observe. And the unrestricted validity of the Schroedinger equation would lead, as also emphasized by Einstein, to the conclusion that the moon, as it is represented in the theory, would be smeared out over the entire night sky. How do we understand this huge disparity between the representation of the universe evolving in accordance with the Schroedinger equation and the empirical reality that we experience?

An adequate physical theory must include a logically coherent explanation of how the mathematical/physical description is connected to the experienced empirical realities. This demands, in the final analysis, a theory of the mind-brain connection: a theory of how our discrete conscious thoughts are connected to the evolving physically described state of the universe, and to our evolving physically described brains.

The micro-macro separation that enters into Copenhagen QM is actually a separation between what is described in quantum mechanical physical terms and what is described in terms of *our experiences*---expressed in terms of our everyday concepts of the physical world, refined by the concepts of classical physics. ([7], Sec. 3.5.)

To pass from *quantum pragmatism* to *quantum ontology* one can treat all *physically described* aspects quantum mechanically, as Von Neumann did. He effectively transformed the Copenhagen pragmatic version of QM into a potentially ontological version by shifting the brains and bodies of the observers---and all other physically described aspects of the theory---into the part described in quantum mechanical language. The entire physically described universe is treated quantum mechanically, and *our knowledge*, and *the process by means of which we acquire our knowledge about the physically described world,* were elevated to essential features of the theory, not merely postponed, or ignored! Thus certain aspects of reality that had been treated superficially in the earlier classical theories---namely "our knowledge" and "the process by means of which we acquire our knowledge"--- were now incorporated into the theory in a detailed way.

Specifically, each acquisition of knowledge was postulated to involve, first, an initiating probing action executed by an "observer", followed by "a choice on the part of nature" of a response to the agent's request (demand) for this particular piece of experientially specified information.

This response on the part of nature is asserted by orthodox quantum mechanics to be controlled by *random chance*, by *a throw of nature's dice*, with the associated probabilities specified purely in terms of physically described properties. These "random" responses create a sequence of collapses of the quantum state of the universe, with the universe created at each stage concordant with the new state of "our knowledge".

If nature's choices conform strictly to these orthodox statistical rules then the retrocausal results reported by Bem cannot be accommodated. However, if nature is not capricious---if God does *not* play dice with the universe---but nature's choices have sufficient reasons, then, given the central role of "our knowledge" in quantum mechanics, it becomes reasonable to consider the possibility that nature's choices are not completely determined in the purely mechanical way specified by the orthodox rules, but can be *biased* away from the orthodox rules in ways that depend upon the character of the knowledge/experiences that these choices are creating. The results reported by Bem can then be explained in simple way, and nature is elevated from a basically physical process to a basically psychophysical process.

The question is then: What sort of biasing will suffice? One possibly adequate answer is a biasing that favors positive experiences and disfavors negative experiences, where positive means pleasing and helpful, and negative means unpleasant and unhelpful.

In classical statistical physics such a biasing of the *statistics* would not produce the appearance of *retrocausation*. But in quantum mechanics it does! The way that the biasing of the quantum statistical rules leads to seemingly "retrocausal" effects will now be explained.

## BACKWARD IN TIME EFFECTS IN QUANTUM MECHANICS

The idea that choices made now can influence what has already happened needs to be clarified, for this idea is, in some basic sense, incompatible with our classical idea of the meaning of time. Yet the empirical results of Wheeler's delayed choice experiments are saying that, in *some* sense, what we choose to investigate now can influence what happened in the past. This backward-in-time aspect of QM is neatly captured by an assertion made in the recent book "The Grand Design" by Hawking and Mlodinow: "We create history by our observations, history does not create us". (p.140)

How can one make rationally coherent sense out of this strange feature of QM?

I believe that the most satisfactory way is to introduce the concept of "process time". This is a "time" that is different from the "Einstein time" of classical deterministic physics. That classical time is the time that is joined to physically described space to give classical Einstein space-time. (See my chapter in "Physics and the Ultimate Significance of Time" SUNY, 1986, Ed. David Ray Griffiths. In this book three physicists, D. Bohm, I. Prigogine, and I set forth basic ideas pertaining to time.)

Orthodox quantum mechanics features the phenomena of collapses (or reductions) of the evolving quantum mechanical state. In orthodox Tomonaga-Schwinger relativistic quantum field theory the quantum state collapses not on an advancing sequence of constant time surfaces (lying at a sequence of times $t(n)$, with $t(n+1)>t(n)$, as in nonrelativistic QM), but rather on an advancing sequence of space-like surfaces sigma(n). (For each n, every point on the spacelike surface sigma(n) is spacelike displaced from every other point on sigma(n), and every point on sigma(n+1) either coincides with a point on sigma(n), or lies in the open future light-cone of some points on sigma(n), but not in the open backward light-cone of any point of sigma(n).)

At each surface sigma(n) a projection operator $P(n)$, or its complement

P'(n)=(I-P(n)), acts to reduce the quantum state to some part of its former self.

For each surface sigma(n) there is a "block universe" defined by extending the quantum state on sigma(n) both forward and backward in time via the unitary time evolution operator generated by the Schroedinger equation. Let the index n that labels the surfaces sigma(n) be called "process time". Then for each instant n of process time a "new history" is defined by the backward-in-time evolution from the newly created state on sigma(n). All predictions about the future are "as if" the future state is the smooth forward continuation from the newly created past. This newly created past is the "effective past", in the sense that the future prediction is given by taking this newly created past to be the past. All empirical traces of the earlier past are eliminated by the quantum collapse.

In orthodox QM each instant of process time corresponds to an "observation": the collapse at process time n reduces the former quantum state to the part of itself that is compatible with the increased knowledge generated by the new observation. This continual re-creation of the effective past is perhaps the strangest feature of orthodox quantum mechanics, and the origin of its other strange features.

The *actual* physical universe is generated by the always-forward-moving creative process. It is forward-moving in the sense that the sequence of surfaces sigma(n) advances into the future. But this forward-moving creative process generates in its wake an associated sequence of revised effective "histories".

Two key features of von Neumann's rules are mathematical formalizations of two basic features of the earlier pragmatic Copenhagen interpretation of Bohr, Heisenberg, Pauli, and Dirac. In association with each observation there is a "choice on the part of the observer" of what aspect of nature will be probed, with an empirically recognizable possible outcome "Yes", and an associated projection operator P(n) that, if it acts on the prior quantum state rho, reduces that prior state to the part of itself compatible with the knowledge gleaned from the experiencing of the specified outcome "Yes".

The process that generates the observer's choice of the probing action is not specified by contemporary quantum mechanics: this choice is, *in this very specific sense*, a "free choice on the part of the experimenter." Once this choice of probing action is made and executed, then, in Dirac's words, there is "a choice on the part of *nature*": nature randomly selects the outcome, "Yes" or "No" in accordance with the statistical rule specified by quantum theory. If nature's choice is "Yes" then P(n) acts on the prior quantum state rho, and if nature's answer is "No" then the complementary projection operator P'(n)=(I-P(n)) acts on the prior state. Multiple-choice observations are accommodated by decomposing the possibility "No" into sub-possibilities "Yes" and "No".

All this is just standard quantum mechanics, elaborated to give a rationally coherent ontological account compatible with the standard computational rules and predictions.

The salient point for us is this. Suppose at some time T (in the past) a system S interacts with a measuring/recording system MR in a way that records in MR the value of a property P(T) of S at time T, whereupon MR moves away from S. And suppose that at time T this property P(T) does not have a well-defined value because the quantum state of S is, say, a 50-50 mixture of two different states with opposite values of P(T). Suppose the state of system S does not evolve after time T, and that a new measurement of the value of property P of S is performed *here and now*, and that some definite outcome, either "Yes" or "No", appears here, according to whether the value of P is positive or negative. Quantum theory then predicts, via the creation of the corresponding new history, an associated reduction of the state of the now-faraway record of the value of earlier state P(T) of S.

The existence of such a correlation is not problematic: it is completely normal and to-be-expected that the two measurement outcomes should be exactly correlated, and that the outcomes in both regions will be 50% "Yes" and 50% "No". But suppose, to illustrate the point with an extreme example, that nature's choice at the later time "now" of its answer to this particular probing question is biased, and delivers the outcome "Yes" 100% of the time. Then there will still be, because of the quantum redefinition of the past, an exact correlation between the two measurement outcomes: both will give "Yes" 100% of the time. Thus the biasing of nature's choice of outcome pertaining to the system S being observed here and now will affect the preserved faraway record of the property P(T) of S at time T: the biasing of the outcome of the observation here and now will shift the result of the observation of the faraway record from 50% "Yes" and 50% "No" to 100% "Yes". The biasing of nature's response here and now has effectively influenced the faraway record of the state of system S at the earlier time T, and influenced also all future predictions that depend upon the state of system S at time T. The biasing of the present choice has altered the *effective* past.

If the question posed here and now about system S were, instead, a different question that nature answers in an unbiased way, then the orthodox rules entail that Nature's (assumed unbiased) faraway choice of a response to the question of pertaining to the recorded measurement of P(T) will be 50% "Yes" and 50% "No". This means that the observer here, by his or her choice of what to measure now, at process time n can send a signal (a sender-controlled message) to the faraway region: the observer's choice of what to observe here and now can influence the *probabilities* of the outcomes of probing actions performed at a later process time n'>n. The concepts of classical relativity theory break down.

It is not so much that the normal history has been altered as that extra *effective* histories have been added, and these extra histories (of the universe) all lead to the favored outcome. Hence the faraway observed record, and all future observations, become altered by the biasing of nature's choice at process time n.

## MATHEMATICAL DETAILS

The description of orthodox quantum mechanics given above is a didactic equation-free account of what follows from the equations of quantum measurement theory. The mathematical details are given in this section.

The mathematical representation of the dynamical process of measurement is expressed by the two basic formulas of quantum measurement theory:

rho(n+1)$_Y$ = P(n+1) rho(n) P(n+1)/Trace (P(n+1) rho(n) P(n+1)),

and

<P(n+1)>$_Y$= Trace (P(n+1) rho(n) P(n+1)) = Trace (P(n+1) rho(n)).

Here the integer "n" identifies an element in the global sequence of probing "measurement" actions. The symbol rho(n) represents the quantum state (density matrix) of the observed physical system (ultimately the entire physically described universe, here assumed closed) immediately *after* the nth measurement action; P(n) is the (projection) operator associated with answer "Yes" to the question posed by the nth measurement action, and P'(n)= (I – P(n)) is analogous projection operator associated in the same way with the answer "No" to that question, with "I" the unit matrix. The formulas have been reduced to their essences by ignoring the unitary evolution *between* measurements, which is governed by the Schroedinger equation.

The expectation value $<P(n+1)>_Y$ is the normal orthodox probability that nature's response to the question associated with P(n+1) will be "Yes", and hence that rho(n+1) will be rho(n+1)$_Y$ . In the second equation I have used the defining property of projection operators, PP=P, and the general property of the trace operator: for any X and Y, Trace XY=Trace YX. (The trace operation is defined by: Trace M= Sum of the diagonal elements of the matrix M).

Consider the familiar example of a pair of systems created in some space-time region, and then traveling to two far-apart labs. The experimenter/observer in each lab chooses to measure some property of the system entering his lab. Let the probing actions in the first and second labs be associated with the projection operators P and Q, respectively.

Suppose you, in your lab, decide to ask whether or not your experience will correspond to the reduction of the current state of the universe (defined by the density matrix rho) to the part of itself, P rho P, compatible with the experience PYes = (The experience associated with the answer "Yes" to the question "Will my experience be the experience associated with the answer "Yes" to the probing action associated with the projection operator P?).

Suppose the observer in the other lab chooses to measure Q. If you know that the other observer is going to measure Q (i.e., is going to see whether nature responds 'Yes' or 'No' to the question "Will I, the second experimenter/observer, experience the thought, feeling, or idea associated with Q?") then how will your knowledge (merely) of what the second experimenter/observer is going to do---or has already done---(namely to choose to measure Q) going to affect your expectations pertaining to what you will see/experience?

The answer is "No Effect! " --- provided the orthodox (pure chance) rules hold.

The point is that the standard prediction in the case that the measurements corresponding to P *and* Q are performed in spacelike separated regions (so that PQ=QP) is that the probability of getting the pair of answers (PYes, QYes) is:

$<PQ>_{YY}$ = Tr (PQ rho)    (Tr rho = 1).

The probabilility of (PYes, QNo) is

$<P, I-Q>_{YN}$ = Tr (P(I-Q) rho) .

Hence your expectation $<P>_{YQ}$ of getting the answer 'Yes' for P if you know (say by prearrangement) that the other experimenter/observer will choose to pose the question corresponding to Q, but have no knowledge of what the other outcome is (was, or will be), but know or believe that the usual statistical (chancy) rules of QM apply, is

$<P>_{YQ}$ = $<PQ>_{YY}$ + $(P(I-Q))>_{YN}$ = Trace (PQ rho) + Trace (P(I-Q) rho)

= Trace (P rho) =  $<P>_Y$.

due to the linearity of the Trace operation.

Thus your expectation, and also the actual probability if the chancy rules really hold, is the same as if the other experiment (corresponding to Q) was not performed, or some different experiment (corresponding to Q1) was performed. This is the standard normal consequence of the chance-based theory: What happens "here" is independent of what is DONE faraway! This is an elementary, but important, consequence of orthodox QM. The normal *statistical* rule entails the normal *causality* rule that what a faraway experimenter freely decides to do "now" cannot affect what you will observe "here and now"!

*Normal causality ideas hold, provided the normal chancy probability rules hold!*

But suppose *nature's choice* of response does not conform to the orthodox statistical rule. Suppose, just to illustrate the main point with an extreme example, that nature's choice is based on certain reasons, and is such that *if* the query corresponding to Q is posed, then nature's answer will definitely be "Yes". Then if the question corresponding to Q is posed, the probability of receiving the answer "Yes" to your local query corresponding to the local operator P will be

$\langle P \rangle'_{YQ}$ = Trace [P ((Q rho Q)/Trace (Q rho Q))] =

Trace (PQ rho)/Trace (Q rho),

where I have again used the projection operator condition QQ=Q, the fact that PQ=QP, and the fact that, for any X and Y, Trace XY= Trace YX.

The matrix ((Q rho Q)/Trace (Q rho Q)) occurring in the above formula is the density matrix that represents the facts that: 1) the original state (of the observed system) is rho; 2) the measurement corresponding to Q is performed; and 3) the outcome is definitely QYes.

In this *biased* case, in which if Q is performed then nature definitely picks outcome QYes, the expectation $\langle P \rangle_{YQ}$ is no longer generally the same as $\langle P \rangle_Y$ = (Trace P rho), which is what it would be if no question were posed faraway. For example, if rho specifies the condition of complete positive correlation of P and Q',

rho= (PQ' +P'Q)/Trace(PQ' + P'Q),

then, from the above result,

$\langle P \rangle'_{YQ}$ = (Trace PQ rho/Trace Q rho )= 0/Trace QP',

which is zero (because PP=P and PP'=0) for the general case in which P, P', Q, and Q' are all nonzero, whereas if no question is posed in the second region, or if the standard chancy rules hold, then the expectation for PYes is

$\langle P \rangle_Y$ = Trace (P rho) = Trace P (PQ' +P'Q)/ Trace(PQ' + P'Q)

= Trace PQ'/ Trace (PQ' + P'Q).

which is not equal to 0, for P and Q' different from zero.

Thus biasing the normal *statistical* rule produces violations of the normal *causality* rule, which asserts that what happens here does not depend upon what is *freely/randomly chosen and done* faraway!

This close interlocking of the normal causality rule with the normal statistical rule is very well known, and was used in my theory of presentiment [7] to predict certain strong presentiment effects, within a quantum framework that allows a biasing of nature's choice of which experience occurs, relative to the normal pure-chance-based rules.

The bottom line is that biasing---relative to the normal orthodox chance-based probabilities---of QYes, for the favored property Q, changes the probability associated with the *other* operator P that is---due to *past* interactions---correlated *in rho* with Q, even though P pertains to events in a region lying now far away from the region where the favored Q was chosen. The free choice of what to measure here now affects the probability for the outcome PYes to occur far away. And the property PYes can be a stored record of what occurred in the intersection of the backward light-cones of the regions in which P and Q are measured. The choice to measure the "favored" property Q---rather than some neutral property Q1---influences (the probability of the stored record P of) what occurred in the past.

This result follows simply from direct application of the orthodox general rules of quantum mechanics, provided the statistical rules can be biased, relative to the normal rules governed by pure chance.

In the Bem experiment with the erotic pictures let $P_{ER}$, $P_{EL}$, $P_{NR}$, and $P_{NL}$ be the projection operators associated with the *observation of the records* of which picture was chosen to appear on which screen, with the subscripts E and N denoting erotic and non-erotic, respectively, and the subscripts R and L identifying right and left screens, respectively. Thus $P_{ER}$ is the projection operator associated with the observable *record* of the fact that the erotic picture was picked and sent to the right-hand screen etc. Let $Q_{ER}$ correspond to the observer-related question "Will I, the subject, see/experience the erotic picture if I open the right-hand screen", etc.. Suppose that the subject has chosen the right-hand screen, and that the PRNG has chosen an erotic picture, and that the RNG has generated a state that has the erotic picture placed in the right-hand screen with probability ½. Then the density matrix rho for the combined PQ system just after the subject has physically interacted with the picture---which has, let us say, then faded away---but before nature's choice of what will appear to the subject is

rho = ($Q_{ER}P_{ER} + Q_{0R}P_{EL}$)/Trace ($Q_{ER}P_{ER} + Q_{0R}P_{EL}$),

where $Q_{0R}$ is operator that corresponds to the subject's looking at the right-hand screen and seeing no picture, and the two terms in the numerator have equal weights when traced. But if nature chooses to actualize experience ER with probability greater than Trace $Q_{ER}$ rho = ½, then one gets the retrocausal effect described above, and reported by Bem: the probability that experience ER will occur is greater than ½, and the probability that the observation of the faraway record associated with $P_{ER}$ will give the answer "Yes" is likewise greater than ½, due to the linkage between $Q_{ER}$ and $P_{ER}$ via their common past. [The process-time machinery helps make clear what happens when the nature's choices are allowed to be biased relative to the normal quantum statistical rules.]

This result is basically a manifestation of the seeming breakdown of normal *causality* concepts if the normal *statistical* rules are not maintained.

This dependence of normal causation upon the validity of the normal quantum rules of chance is, of course, well known! But the upshot here is the pertinent conclusion that making the dynamics *more rational*---by saying that nature's choices have reasons, and are thus not purely random---makes the dynamics *less causal*.

The failure of normal ideas about causation is achieved not by foisting some irrational or unnatural ad hoc condition on the dynamics, but rather by merely rationally insisting that the choices made by nature *stem from sufficient reasons, and that two such reasons are to promote the positive and to curtail the negative experiences of observers.*

Such a proposal goes against what some scientists believe to be the proper duty of science and scientists, namely to refute any such idea of the non-neutrality of nature in favor of the idea that nature maintains strict neutrality by playing dice with the universe--Einstein notwithstanding.

Science has certainly made great strides in reducing any clear need for a biasing of nature's inputs into the physical dynamics. It is certainly worthwhile to pursue efforts to eliminate any empirically-mandated need for nature's input into the quantum dynamics to be nonrandom, but not to the extent of banning from publication in scientific journals seemingly high quality reports of empirical results that appear to contradict the object of those endeavors. A scientific theory is immune to scientific falsification if the only evidence accepted in scientific discourse is data compatible with the current theory.

If natural process does indeed involve a biasing of nature's choices, then it may be of great practical importance for us to explore by the tools of science the details of the structure of this biasing, not merely to satisfy our idle curiosity, but also to allow us to use the thus-expanded science-based knowledge of the workings of nature to improve the quality of our lives.

In Bem's Experiment 2, "Precognitive Avoidance of (Subliminal) Negative Stimuli", a long sequence of pairs of pictures is shown to the subject, who chooses a 'preferred' picture from each pair. After each such choice, an RNG makes a random 50-50 choice of which picture in the pair to identify as the 'target', and then flashes a positive or negative subliminal picture according to whether this target is the preferred or un-preferred picture. If the targeting and associated subliminal stimulus were to occur *before* the subject's choice of preferred picture then there would be no problem with empirical evidence that the positive and negative *prior* stimuli appear to influence the subject's *later* choice of preference. But the normal idea of forward causation does not allow the targeting and associated stimulations that occur *after* the choice of preference to affect that earlier choice of preference. Yet Bem's result is that the pictures chosen *later* as targets are more likely to be *preferred earlier*, compared to the pictures chosen later to be non-targets.

This apparently retrocausal effect flows automatically from the assumption that nature's choice of which outcome actually occurs has a tendency to accentuate the positive and eliminate the negative: such a putative biasing has the effect of adding to the density matrix rho, after nature's biased choice, an abnormal term that corresponds to the *addition* of extra effective histories that lead to positive feelings, or to the *elimination* of effective histories that lead to negative feelings. These extra effective histories have essentially the same effect on the background state of the subject's brain during the process of his or her choice of pleasure versus nonpleasure as the normal effect of performing the brain-changing action *before* the subject's choice of response. In both cases the effective state of the brain of the subject during his or her process of choosing is changed in essentially the same way: whether the change in the effective underlying background state of the subject's brain comes from changes in earlier or later boundary conditions is not important.

To achieve this explanation it must be recognized that the von Neumann process-1 action of posing a question identifies a complex property of the brain of the observing system that can be grasped as a meaningful idea-representing whole: true consciousness is not demanded.  Such a subconscious form of process-1 action is probably needed also to provide a rationally coherent account of psychogenesis, from pre-biological times to the age of homo sapiens.

Bem's experiment 3 is "Retrocausal priming I". A sequence of pictures is shown to the subject, who responds to each picture by pressing a first or second button according to whether he or she feels the picture to be pleasing or not. After the response of the subject to the picture is recorded, a 'word' is selected by a PRNG, and shown to the subject. This word will be either 'congruent' or 'incongruent' with the picture: it will have either the same or opposite positive or negative valence. Bem reports that the recorded time that it takes for the subject to respond to the question of whether or not the picture is 'pleasing' is longer or shorter according to whether the *subsequently* chosen and displayed 'word' is congruent or not with the previously displayed picture. Again the question is: How can the *recorded facts* about what occurred *earlier*---in particular the response time---depend on which word was *later* randomly selected and shown to the subject?

The answer is the same as before: the effective background state---positive or negative---in the extra effective histories created by nature's biased response to the later priming is similar to the normal forward-in-time effect of the same priming: it matters little whether this influential background state is produced by an initial or a final boundary condition.

The same general argument carries over to all nine of Bem's experiments. Of course, the strength of the effect depends upon the power of the subject's conscious or subconscious states to affect nature's choices. This power appears to vary among subjects.

# WHY DO THE NORMAL STATISTICAL RULES NORMALLY HOLD?

Suppose that the Principle of Sufficient Reason does hold, so that each of nature's choices has a reason to be what it turns out to be. And suppose that these reasons lead to choices that violate the orthodox statistical rules. Then violations of normal ideas about causation may occur. But the question then arises: Why do the normal orthodox quantum rules work as well as they do?

The answer is that we considered above an extreme case in which there was a connection between P and an *identified suspected cause Q*. Normally there can be many entangled Q's that could enter into nature's sufficient reasons, and the favored-by-nature relevant variable Q (in our special case associated with the human subject's emotional experience) will generally be unknowable to the observers of P. In general the scientist will have no idea of which features of the world are driving nature's choices in a given actual situation. In these usual cases the scientist must perform an averaging that reflects his ignorance.

The usual *classical* way to represent a complete lack of knowledge about the variables in some domain is to average over the range of variables in that domain, ascribing equal weights to equal volumes of phase space. This is the weighting that is invariant under canonical transformations. The *quantum* analog is to take the Trace, which is invariant under unitary transformations, over the domain of factors about which we have no knowledge.

A complete lack of knowledge about the identity of Q, means that we should average Q over the whole set of Q's unitarily equivalent to it, within the full space in which it lies (which is a component of a tensor product of spaces), and about which we lack knowledge. But this averaging, needed to account for the lack of knowledge about what reasons are driving nature's choices, will effectively erase all dependence on the variables about which one has no knowledge, and reduce the rule for computing expected probabilities to the usual quantum mechanical rules associated with the notion of pure chance.

In more detail the point is this. If nature's choice has a reason, and this reason impels it to answer "Yes" to the posed question corresponding to Q, then the expectation $<P>'_Q$ of P, given that Q is performed and that nature's answer to the Q question is "Yes", is

$<P>'_Q$ = Trace [P ((Q rho Q)/Trace (Q rho Q))]

   = Trace PQ rho/Trace Q rho

as already discussed. But suppose that Q is not known. Suppose, for example, that the various possible Q are identified by points on a circle, labeled by the angle θ, and that every point θ on the circle has equal a priori weight. Then the expectation of P is

$<P>'_Q$ = Trace P(1/2π)∫dθ Q(θ) rho /Trace (1/2π)∫dθ Q(θ) rho.

One must integrate over the unknown variable, assigning equal a priori weights to each possibility.

In our case this example generalizes to

$<P>'_Q$=  [Trace (P Integral over all U of  UQU` rho)]/
       [Trace (Integral over all U  of  UQU' rho)]

   = Trace (P rho)/Trace (rho)

where U`is the Hermitian conjugate of U, and the integral is over the invariant Haar measure on the (compact) space of unitary matrices, and I have used the fact that

(Sum over all U of  U Q U` ) = I (Trace Q/Trace I),

where Q is a projection operator, I is the unit matrix/operator, U`  is the Hermitian Conjugate of U, and the sum over U is a sum over all unitary matrices U, with weights specified by the normalized invariant Haar measure, which is mapped invariantly onto itself under any unitary transformation from either left or right.

This result means that if Q is unknown then the probability <P> is just what is given by the usual pragmatic statistical rule, which, however, now arises from choices at *the basic ontological level that accord with the principle of sufficient reason.*

## CONNECTION TO WHEELER'S DELAYED CHOICE

The Bem backward-in-time effects can be viewed as a corollary of a well-known backward-in-time property of QM, namely the Wheeler "delayed choice" effect, combined with a *biasing*---i.e., a violation---of the normal chancy weightings of nature's choices of responses to probing actions pertaining to emotional states.

As regards the standard Wheeler delayed-choice effect itself, the *orthodox* theory entails that when nature makes her choice of response, the past is 'effectively reduced' to the portion of the former past that fits smoothly onto the new, reduced, state of the universe, which nature has just chosen. The *parts of the former past* that conflict with nature's current choice are *effectively* eliminated. Here "effectively" means "for the purpose of making predictions pertaining to the future": As far as the potentialities for the next event are concerned, it is just "as if" the past were now "reduced" to that part of the former past that evolves into the new contemporary reality, created by nature's current response, with the remainder of the former past suddenly eradicated.

[Wheeler's delayed choice experiment is essentially this: Suppose, during a double-slit experiment, at a time Tf before a photon reaches your eye, but after the photon has passed through the slits, you focus your vision on the slits through which photons that are coming, one at a time. Then you will "see" that each photon passed at the earlier time *T*'---prior to your choice of how to focus your eyes---through one slit or the other, not both. But if at the later time Tf you choose to focus your eyes straight ahead then you will see the particles building up a pattern of stripes that depend upon the distance between the two slits, indicating that the wave packet went, at the earlier time *T*, through both slits: the later choice at time *Tf* on the part of the observer of what to observe influences the content of the *effective* quantum state at the earlier time *T*. This redefinition causes no conceptual problem in the orthodox theory because the physically described quantum state is not a material reality: it is merely a representation of *potentialities for future* *experiences of observers*, and each of these experiences depend upon what the observer chooses to observe.

The *actual* evolution proceeds in a well-ordered sequence, with each event associated with a finite (small) space-time region of zero temporal thickness, no part of which lies in the backward light-cone of any point in any of the regions associated with any earlier (in the ordered sequence of events) event. (See [9], Fig. 13.1) Each event creates a new *effective* past, but does not alter any past actual event.

The notorious "nonlocality" feature of orthodox quantum mechanics can be attributed to this "delayed choice" effect of nature's present choice upon the new effective past. This new effective past, *created by the prolongation of the newly created present physical state into the past via the (inverse of) the Schroedinger equation,* is only a portion of what was formerly present. The effective elimination of parts of the former past *effectively eradicates the records of the parts of the past that have been eliminated.* Thus the reduction of the state rho associated with the measurement made *here and now* can affect the potentialities associated with faraway observations of records pertaining to what led up to the measurement made here and now. These retrocausal effects become, in this way, no longer a vague mysterious phenomena, but rather an understandable consequence of the elimination of pure chance from the basic quantum psychophysical laws of nature.

## CONCLUSION

Numerous reported seemingly backward-in-time causal effects are naturally explainable within forward-in-time orthodox quantum mechanics, provided the orthodox input of pure chance is replaced by the input of sufficient reason, with two such reason being the promotion of positive, and the suppression of negative, feelings, or their subconscious relatives generated by subliminal stimuli.